\documentclass[12pt,a4paper]{article}
\usepackage{latexsym}
\usepackage[latin1]{inputenc}
\usepackage[dvips]{graphics}
\usepackage{amssymb,amstext,amsfonts,amsmath}
\begin{document}
%\maketitle

%\newpage

% PACS Combinatorics, 02.10.Ox, statistical mechanics quantum, 05.30.-d

%\tableofcontents % SPELLLLL CHECKEEEDDDD!!!!!!!!!!!!

\begin{center}
{\Large  Construction of density operator for a general mean-field Hamiltonian and its application to the models of correlated fermions \\}
\vspace{0.5cm}
%\large{
Jakub J\c{e}drak$^{\ast}$ and Jozef Spa\l ek$^{\dag}$  \\ %$^\ast$ \\ %} \\
Marian Smoluchowski Institute of Physics, Jagiellonian University, \\
Reymonta 4, 30-059 Krak\'{o}w, Poland \\
\end{center}
%\vspace{0.8cm}

%\newpage
\begin{abstract}

We analyze a class of mean-field (MF) lattice-fermion Hamiltonians and  
 construct the corresponding grand-canonical density operator for such system.
New terms are introduced, which may be interpreted as local fugacities, molecular fields, etc. 
The presence of such terms is an unavoidable consequence of a consistent statistical description. 
Although in some  cases (e.g. in the Hartree or the Hartree-Fock-type approximation) the presented formalism  is redundant, in general (e.g. for a renormalized t-J model) it leads to nontrivial modifications of the thermodynamic properties. 
%The application of such obtained density operators as trial states for some variational principles is discussed.
%The case of zero temperature is also briefly analyzed. 
\end{abstract}
% above mentioned
%\vspace{1.0cm}
% application of
%% PACS 71.27.+a 	Strongly correlated electron systems; heavy fermions
%% statistical mechanics,  quantum, 05.30.-d
%71.10.Fd, 02.10.Ox,
% i.e. depending on $\mathbf{C}$-numbers,  having  a meaning of the expectation/ mean values of some operators.

%%%%%%%%%%%%%%% I N T R O %%%%%%%%%%%%%%%%%%%%%%%%%%%%%

\section{Introduction}

For most of  realistic models of quantum many-particle systems an exact solution cannot be achieved. This is a typical situation condensed matter or  nuclear physics. As a consequence, a number of approximate methods have been developed. Among them, widely used are various types of the mean-field approaches.

By mean-field (MF) description of the problem we understand  making use of the Hamiltonian which depends on  extra parameters, having the meaning of expectation values of well-defined operators. Values of those parameters  are not   \textit{a priori} known and are to be determined.  The  Hartree-Fock method is a good  example of such approach.   %, and the detailed form of GMF Hamiltonian may be   either result of the replacement of some operators in some other original/ precursor Hamiltonian, or postulated ...
  
MF methods provide us with a valuable insight into the physical properties of a system under the consideration. They  are capable of describing many interesting phenomena, especially those involving phase transitions,  such as that from normal metal to  superconductor transition (e.g. in the form of the Bardeen-Cooper-Schrieffer theory), or those from a paramagnet to either antiferro- or ferromagnet.  

In implementing such a procedure, it is the  physics of the problem, not the mathematics, that tells us when the application of MF approach is justifiable.  Nonetheless,  we  may also be faced with the problem of considering the internal mathematical consistency of the method, as well as its underlying general principles. The aim of this paper is to clarify this issue to some extent, in a simple manner.%/to put some light on this issue.

The paper is organized as follows. Sec.  \ref{statement of the} contains  formulation of the problem and invokes basic relevant notions from statistical mechanics. In Sec. \ref{meanfield}, by means of the maximum entropy principle we construct the statistical-mechanical description of the system in the  mean-field case.  Explicitly, in \ref{consturction} we derive the grand-canonical (GC) MF density operator, in \ref{Thermodynamics} we comment on the MF thermodynamics and its consistence with the MF statistical mechanics, whereas in \ref{Hartree} we discuss the important case of the Hartree-Fock (HF) type MF Hamiltonians. Finally, in \ref{Bogoliubov} the connection of present method with the variational principle based on the Bogoliubov inequality is elaborated. In Sec. \ref{examples} two specific  examples of mean-field Hamiltonians are provided, for which our approach leads to significant corrections. Section \ref{summary} contains a summary. % In Appendix A we provide some of the technical details of our analysis. 
 
%, both  where the greater care is necessary, and where our results are not redundant.
%Number of exactly solvable models is very limited. 
%%%%%%%%%%%%%  S T A T E M E N T
\section{Statement of the problem\label{statement of the}}
We assume that the Hilbert-Fock state space is of finite dimension $D$. The MF Hamiltonians are denoted as $\hat{H}$, whereas those  which are not of MF type, will be called \textit{exact} and labelled with the subscript '\textit{e}'. 
%We assume also the  applicability of the equilibrium (quantum) statistical mechanics. 

As already mentioned in the Introduction, by  \textit{mean-field approach} we understand the situation in which Hamiltonian depends on certain parameters, that are to be determined when  solving the problem at hand. %Let $\hat{H}$ depend on $M$  such parameters, $A_1, A_2, \ldots,  A_M$.
%(this dependence may be not denoted explicitly).  $\hat{H}$ 
Such  MF Hamiltonian is usually derived from the exact Hamiltonian $\hat{H}_{e}$, by replacing some of the operators, $\{ \hat{A}_s \} = \hat{A}_1, \ldots, \hat{A}_M$,   appearing in $\hat{H}_{e}$,   by  $ \mathbb C$-numbers $\{A_s \} = A_1, A_2, \ldots,  A_M$ $(\equiv \vec{A})$. 
Those numbers have a natural interpretation of expectation values of the corresponding operators $\{\hat{A}_s\}$. However, the mean-field Hamiltonian  $\hat{H}$ may be not be related in an obvious manner to any exact Hamiltonian.  
 
We assume that the system is in thermal equilibrium with the particle reservoir characterized by the temperature %\footnote{We put Boltzmann constant equal to unity, $k_{B}=1$.}%\footnote{In principle, one can have $Q$ different species of fermions, then we have obviously $Q$ corresponding chemical potentials. For simplicity we assume from now on, that only one kind of particles is present.} 
 $T = \beta^{-1}$  and the chemical potential $\mu$.  The  question then arises, how to determine the values of  parameters $A_s$  appearing in $\hat{H}$. 
If the Hamiltonian of the system were exact, i.e. not expectation-value dependent, then the treatment of such situation is straightforward, in the sense, that the  grand canonical (GC) ensemble is well defined, and all information about the system is contained in the  density operator \cite{statistical mechanics, Zubariev} defined as
\begin{equation} 
\hat{\rho}_{e} = Z_{e}^{-1} \exp \big(-\beta ( \hat{H}_{e} - \mu  \hat{N})\big), 
\label{exact density matrix GCE}
\end{equation}
%where $ex$ means \textit{exact}, i.e.  not  expectation-value-depended, operator.
where $Z_{e} =  \text{Tr}[  \exp \big(-\beta ( \hat{H}_{e} - \mu  \hat{N})\big) ] $ is the grand-canonical partition function. The expectation value of any observable $\hat{A}$ is then given by
\begin{equation}
\langle \hat{A}\rangle_{e}  \equiv  \text{Tr}[\hat{A}  \hat{\rho}_{e}].
\label{mean value of A exact}
\end{equation}
Moreover, we assume, that in the mean-field case  this basic definition  still holds, i.e.
\begin{equation}
\langle \hat{A}\rangle  =  \text{Tr}[\hat{A}  \hat{\rho}(\vec{A})].
\label{mean value of A}
\end{equation}
Regardless of the  form of  MF  density operator $\hat{\rho}$,  we should be able to obtain any expectation value through (\ref{mean value of A}). In particular, the last definition allows us to determine the values of $A_1, A_2, \ldots,  A_M$ appearing in $\hat{\rho}$, 
by solving  a system of  $M$ implicit  equations for $M$ variables (the so-called self-consistency equations), which take the form 
\begin{equation}
\langle \hat{A}_{s}\rangle  = {A}_{s}  = \text{Tr}[\hat{A}_{s}  \hat{\rho}(A_1, A_2, \ldots,  A_M)], ~~~~ \forall_{s}.
\label{mean value of A set of eq}
\end{equation}
We may hope that this system should  have at least one  solution.  % G_{s}(A_1, A_2, \ldots,  A_M)

One would expect, that MF grand-canonical density operator should be given by (\ref{exact density matrix GCE}) with $\hat{H}_{e}$ replaced with $\hat{H}$, so
\begin{equation}
\hat{\tilde{\rho}} = \tilde{Z}^{-1} \exp \big(-\beta ( \hat{H} - \mu  \hat{N})\big), ~~~~\tilde{Z} =  \text{Tr}[  \exp \big(-\beta ( \hat{H} - \mu  \hat{N})\big) ]. 
\label{MF naive DO}
\end{equation}
This is, however, not necessarily the case, as is discussed in detail below.
%%%%%%%%%%%%%%%%%% below I make changes according to Boss's remarks!
In order to find the correct mean-field GC density operator for a given MF Hamiltonian $\hat{H}$, we  proceed  analogously to the exact (non-MF) case and \textit{derive} the form of such density operator from the basic principles of statistical mechanics.
Thus, before concentrating in the remaining part of the paper on the mean-field case, we first  would like to summarize briefly the non-MF situation.
Namely, for the exact Hamiltonian $\hat{H}_{e}$, the GC density operator (\ref{exact density matrix GCE})  may be derived  from the  requirement that it  maximizes the  von Neumann entropy, $S \equiv - \text{Tr} \hat{\rho}_{e} \ln \hat{\rho}_{e}$, under the conditions that the expectation values of $\hat{H}_{e}$ and  of the particle number operator $\hat{N}$, are fixed. This means, that we maximize the following functional (see \cite{statistical mechanics, Zubariev})
\begin{equation}
\mathcal{S}_{e} =   \text{Tr} \big( -\hat{\rho}_{e} \ln \hat{\rho}_{e} -\beta  \hat{\rho}_{e} \hat{H}_{e}
+ \beta \mu  \hat{\rho}_{e} \hat{N} - \omega \hat{\rho}_{e} \big ).
\label{entropy exact GCE}
\end{equation}
The first two conditions are enforced with the help of the Lagrange multipliers $\beta$ and $\mu$, whereas the parameter $\omega$ is introduced to ensure the normalization, $\text{Tr} [\hat{\rho}_{e}]=1$. 
Furthermore,  $\hat{\rho}_{e}$ must be a  constant of motion, so from the quantum Liouville equation \cite{statistical mechanics, Zubariev}
\begin{equation}
i \hbar  \frac{\partial}{\partial t}  \hat{\rho}_{e}(t) = [ \hat{H}_{e}, \hat{\rho}_{e}(t)],
\label{quantum Liouville}
\end{equation}
it follows, that $ [ \hat{H}_{e},  \hat{\rho}_{e}] = 0$. If, additionally, $ [ \hat{H}_{e},  \hat{N}] = 0$, which is usually the case in non-relativistic quantum mechanics, we may simultaneously diagonalize each of the operators appearing in (\ref{entropy exact GCE}).

Parenthetically, the maximization procedure is most conveniently carried out in the basis, in which $\hat{\rho}_{e}$ is  diagonal, i.e. $\hat{\rho}_{e} = \sum_{i}^{D} p_i |i \rangle \langle i|$.  % w_1, w_2, \ldots, w_D \}.
The resulting $\hat{\rho}_{e}$ is then of the form given by (\ref{exact density matrix GCE}), \cite{Zubariev}. Density operators for the canonical and the micro-canonical ensembles may be obtained in a similar manner. 
The fact, that those density operators maximize the entropy functional (\ref{entropy exact GCE}) or analogous expressions, may be considered as their basic property  \cite{Jaynes}.

%\footnote{Formally, (\ref{entropy exact GCE}) is a functional, and the derivatives should be in principle understood as the functional ones. However, for finite-dimensional Hilbert space, and with  no  dependence of  $\hat{\rho}_{e}$ on time, technically the problem reduces to maximization of the ordinary function of many variables.}
%Such a non-standard situation is encounter when one deals with a generalized mean-field Hamiltonians with / of arbitrary dependence on $\vec{A}$. %%%%%%%%%%%%%%%
%%%%%%%%%%  E N T R O P Y  M F
\section{Maximum entropy principle in the mean-field case \label{meanfield}}
\subsection{Construction of the mean-field density operator \label{consturction}}
We turn now to the  construction  of the proper form of MF density operator. We concentrate on the case, for which the entire system description is based \textit{solely} on  MF Hamiltonian.  In the previous Section it was stated, that maximum entropy condition  is basic for canonical density operators. Moreover, the principle of maximum entropy is one of basic postulates of statistics, it ensures uniqueness and consistency of probability assignments, \cite{Jaynes}.  Consequently, it should be applied also to the particular case of mean-field description. 
Now, the difference with the standard, i.e. non-MF situation above is that the Hamiltonian depends on $M$ additional variables $\{A_s\}\equiv \vec{A}$, interpreted as the expectation values of corresponding operators $\{\hat{A}_{s}\}$. %,  {mean value of A set of eq} see defined in a standard way with the help of equation (\ref{mean value of A}). 
Those $M$ expectation values, together with $D$ diagonal matrix elements of $ \hat{\rho}$, are now the variational parameters, with respect to which the functional%\footnote{Although we use the term \textit{entropy} for the two distinct quantities, the von Neumann entropy $S$ and the just introduced entropy functional $\mathcal{S}$, we hope there is no danger of confusion.}
\begin{equation}
\mathcal{S} =  \text{Tr} \big( - \hat{\rho} \ln  \hat{\rho} -\beta  \hat{\rho} \hat{H}(\vec{A})
+ \beta \mu \hat{\rho} \hat{N} - \omega (\hat{\rho} - \frac{1}{D}) \big),
\label{entropy MF no lambdas}
\end{equation} 
is maximized. As earlier, $ - \omega (\hat{\rho}  - 1/D)$ term\footnote{Here taken in the form which slightly differs from that of (\ref{entropy exact GCE}), but this technical alteration is inessential.} is added to ensure normalization of the density operator. Next, if those  $M+D$  parameters  were treated as  independent, there is no certainty that $\{A_s\}$ obtained in that manner coincide with those resulting form the self-consistency conditions  (\ref{mean value of A set of eq}).   
In principle, one may use (\ref{mean value of A set of eq})
to eliminate the $A_s$ variables from $\hat{H}$.  Nonetheless, the resulting dependence of $\mathcal{S}$ on $D$ diagonal elements of $\hat{\rho}$ is usually  very  complicated, and the MF density operator cannot be obtained in an explicit form.  

To resolve the situation, we use the  Lagrange-multiplier method. Explicitly, the $M$  conditions (\ref{mean value of A set of eq}) are  enforced with the help of $M$ Lagrange multipliers $ \{\lambda_{s} \}\equiv \vec{\lambda} $. To  account for that, the terms of the form $\lambda_{s} (\text{Tr}[ \hat{\rho}\hat{A}_{s}] - A_{s} ) $ must be added to the functional (\ref{entropy MF no lambdas}). Under these circumstances, we have in total $D + 2M $ variables, $\{p_1, \ldots, p_D, A_1, \ldots, A_M,  \lambda_1, \ldots, \lambda_M \}$, treated as independent (apart from the normalization condition). In effect, the constrained entropy functional reads 
\begin{equation}
\mathcal{S}_{\lambda} =  \text{Tr} \Big[ - \hat{\rho}_{\lambda} \ln  \hat{\rho}_{\lambda} -\beta \big( \hat{\rho}_{\lambda} \hat{H}
-  \mu \hat{\rho}_{\lambda} \hat{N} - \sum_{s=1}^{M}  \lambda_{s}\hat{\rho}_{\lambda} (\hat{A}_{s} -  A_{s})\big) - \omega (\hat{\rho}_{\lambda} - \frac{1}{D})\Big].
\label{entropy MF}
\end{equation} 
The incorporation of the self-consistency constraints may be regarded as a redefinition of the MF Hamiltonian
\begin{equation}
\hat{H} ~~\to ~~ \hat{H}_{\lambda} ~= ~  \hat{H}-\sum_{s=1}^{M}   \lambda_{s}(\hat{A}_{s} - A_{s} ).
\label{GMF Hamiltonian with lambdas}
\end{equation}
The last step  is the basic novel ingredient of our method  and is discussed in detail in the remaining part of this paper.

Few  remarks are in place  here. First, we are mainly interested in a situation when  each operator $\hat{A}_{s}$ is bilinear in creation and/or  annihilation operators\footnote{In principle, the presented formalism applies also to the  general situation of the arbitrary mean-field Hamiltonian.}, as then $\hat{H}_{\lambda}$ is easily tractable. Second, some  combination of  the  operators  $\{ \hat{A}_{s}\}$ appearing in (\ref{GMF Hamiltonian with lambdas}) may be proportional to the number operator $\hat{N}$. Nonetheless, the presence of the chemical potential $\mu$ fixes the  average number of particles $N$. % This is the reason for using the MF approach combined with grand canonical ensemble formalism in such case, otherwise;  we will not be able to specify $N$. 

Third, the   operators $\hat{H}$ and $\hat{N}$ in (\ref{entropy MF})  have a different  status than the operators $\{ \hat{A}_{s}\}$.  The expectation values of the former are assumed to be fixed \textit{a priori}, enforced by the values of  temperature and chemical potential. In contrast, the expectation values of the latter are obtained from a variational procedure, and are not \textit{a priori} given.
Consequently, the same distinction holds for parameters  $\beta$, $\mu $ from one side and  $\{ \lambda_{s}\}$ from the other, even though all of them play the role of Lagrange multipliers. Nonetheless, some of  $\{ \lambda_{s}\}$ may have a physical interpretation. Depending on the corresponding  operator $ \hat{A}_{s}$, they may be termed as local fugacities, molecular fields, etc. Note, that for $ \hat{A}_{s} =   \hat{A}^{\dag}_{s}$,  $ \lambda_{s}$ is real, but for non-Hermitian $ \hat{A}_{s}$, it is a complex number in general. % For some $ \hat{A}_{s}$, $ \lambda_{s}$ the physical interpretation is, however, not that obvious.

Apart from $ \hat{H}$ and  $\hat{N}$, there may be more observables with expectation values fixed \textit{a priori}. In such a situation the  terms $(- \xi_s \hat{A}_s)$, analogous to $- \mu \hat{N}$ term, should be also added, to   ensure that the quantity $\langle \hat{A}_s  \rangle$ is consistent with the prior information about the system.  If we ignore some available information about the physical system under consideration, we obtain a broader probability distribution.

Fourth, the operators appearing in  (\ref{GMF Hamiltonian with lambdas}) do not necessarily  commute with each other, and thus cannot be, in general, simultaneously diagonalized. However, in analogy to the canonical case, we require that $\hat{\rho}_{\lambda}$ thus obtained is stationary, i.e.  $  \partial   \hat{\rho}_{\lambda}(t)/\partial t =0$, from which, using the quantum Liouville equation (\ref{quantum Liouville}) for the MF case the condition $[\hat{H}_{\lambda}, \hat{\rho}_{\lambda}] = 0$ follows.
In particular, in the case  $[\hat{N}, \hat{H}_{\lambda}] \neq 0$ (e.g. in the BCS theory  of superconductivity), we can also redefine the Hamiltonian according to $\hat{H}_{\lambda} \to \hat{K}_{\lambda} = \hat{H}_{\lambda} - \mu \hat{N}$, and then demand, that the corresponding grand Hamiltonian commutes with the density operator, $[\hat{K}_{\lambda}, \hat{\rho}_{\lambda}] = 0$. In result,  diagonalization of $\hat{K}_{\lambda}$  provides us again with  the set  of $\vec{A}$  and $\vec{\lambda}$- dependent eigenvectors $\{ | i(\vec{A}, \vec{\lambda})\rangle \}_{i=1}^D$ and corresponding eigenvalues of the density operator $ \hat{\rho}_{\lambda}$. 

Fifth, the present formalism may be easily modified and applied to the case of canonical ensemble or discrete classical models (e.g. lattice gas, or spin systems).  

Finally, even if, strictly speaking, the $T=0$ situation cannot be investigated directly within the framework of the present formalism, (must be analyzed separately), it is fully legitimate as well as convenient to analyze the situation in question  as the $T\to 0$ limit, e.g. for finite, but sufficiently low temperature. A more detailed analysis of this point will be presented elsewhere \cite{Jedrak}.
 
%\subsection{Maximization of the constrained entropy in the MF case}
Let us go back to the problem of finding the maximum of the entropy subject to the constraints. 
The functional (\ref{entropy MF}), written in the common eigenbasis of $\hat{K}_{\lambda}$ and $\hat{\rho}_{\lambda}$ reads 
\begin{equation}
\mathcal{S}_{\lambda} =  \sum_{i=1}^{D} \Big\{ - q_{i} \ln  q_{i} -\beta  q_{i}\Big( (\hat{H})_{ii} 
-\mu  (\hat{N})_{ii}  - \sum_{s=1}^{M}    \lambda_{s}  \big( (\hat{A}_{s})_{ii} -  A_{s}\big)\Big)  - \omega (q_{i} - \frac{1}{D}) \Big\},
\label{entropy MF eigen}
\end{equation} 
(with $ (\hat{N})_{ii} = \langle i | \hat{N} | i \rangle $ etc.) We assume, that the maximum of (\ref{entropy MF eigen}) corresponds to stationary point. Then, the necessary conditions for an extremum of $\mathcal{S}_{\lambda}$ with the normalization preserved, are 
%%%%%%%%%%%%%
\begin{equation}
\frac{\partial \mathcal{S}_{\lambda}}{\partial  \omega}  =  1 - \sum_{j=1}^{D}  q_j = 0,
\label{derivative omega}
\end{equation}

\begin{equation}
\forall_{j}:  ~~~ \frac{\partial \mathcal{S}_{\lambda}}{\partial  q_{j}}  = - (1 + \omega) - \ln q_j -\beta  (\hat{H})_{jj}  + \beta \mu  (\hat{N})_{jj}  + \sum_{s=1}^{M}   \beta\lambda_{s}  \big( (\hat{A}_{s})_{jj} -  A_{s} \big)  =   0,
\label{derivative q i}
\end{equation}
%%%%%%%%%%
\begin{equation}
\forall_{w}:~~~   \frac{\partial \mathcal{S}_{\lambda} }{\partial  A_{w}}  =  -\beta \sum_{i=1}^{D} 
q_{i}   \Big(\frac{\partial \hat{H}_{ii} }{\partial  A_{w}}   + \lambda_{w} \Big) = 0,
\label{derivative A s}
\end{equation}
and
\begin{equation}
\forall_{w}:~~~ \frac{\partial \mathcal{S}_{\lambda}}{\partial  \lambda_{w}} =  \beta \sum_{i=1}^{D} q_i  \big( (\hat{A}_{w})_{ii} -  A_{w})  \big)  = 0.
\label{derivative lambda s}
\end{equation}
In (\ref{derivative A s}) and (\ref{derivative lambda s}), we ignored the possible explicit $\vec{A}$- and $\vec{\lambda}$- dependence of the (normalized) eigenvectors $|i \rangle$. This is justified due to the  Hellmann-Feynman theorem, \cite{Feynman theorem}. Using (\ref{derivative omega}) and (\ref{derivative q i})   we obtain the explicit, basis-independent form of  density operator $\hat{\rho}_{\lambda}$, 
\begin{equation} 
 \hat{\rho}_{\lambda} =  \mathcal{Z}_{\lambda}^{-1} \exp \big(-\beta ( \hat{H}_{\lambda} - \mu  \hat{N})\big), ~~~  \mathcal{Z}_{\lambda}= \text{Tr}[\exp \big(-\beta ( \hat{H}_{\lambda} - \mu  \hat{N})\big)].
\label{correct MF density matrix}
\end{equation}
We see that (\ref{correct MF density matrix}) differs from (\ref{MF naive DO})  by the presence of additional terms in the Hamiltonian, in accordance with (\ref{GMF Hamiltonian with lambdas}). Additionally, Eqs. (\ref{derivative A s}) and (\ref{derivative lambda s}) may be rewritten respectively as 
\begin{equation}
 -\frac{1}{\beta} \frac{\partial \mathcal{S}_{\lambda} }{\partial  A_{w}}  = \text{Tr}[ \hat{\rho}_{\lambda}\Big(  \frac{\partial  \hat{H} }{\partial A_w}  + \lambda_w  \Big)] = \Big \langle  \frac{\partial  \hat{H} }{\partial A_w} \Big \rangle_{\lambda}  + \lambda_w   =      \Big  \langle   \frac{\partial  \hat{H}_ {\lambda} }{\partial A_w} \Big  \rangle_{\lambda} = 0,  
\label{mean value of H GMF with lambda, d/dA}
\end{equation}
and
\begin{equation}
\frac{1}{\beta} \frac{\partial \mathcal{S}_{\lambda} }{\partial  \lambda_{w}}  = \text{Tr}[ \hat{\rho}_{\lambda}(\hat{A}_w -  A_{w})]= 
\big \langle  \hat{A}_{w} \big \rangle_{\lambda}  -  A_{w} = \Big \langle   \frac{\partial  \hat{H}_ {\lambda} }{\partial \lambda_w} \Big \rangle_{\lambda} = 0.   
\label{mean value of H GMF with lambda, d/dlambda}
\end{equation}
Next, we define the generalized  grand-canonical Landau functional 
\begin{equation}
\mathcal{F}(\vec{A}, \vec{\lambda}) \equiv -\beta^{-1}\ln \mathcal{Z}_{\lambda} = -\beta^{-1} \mathcal{S}_{\lambda}( \hat{\rho}_{\lambda}(\vec{A}, \vec{\lambda}), \vec{A}, \vec{\lambda}),  
\label{mathcal F functional}
\end{equation}
with $\mathcal{Z}_{\lambda}$ given by (\ref{correct MF density matrix}). That last relation may be easily verified by inserting $\hat{\rho}_{\lambda}$ given by 
(\ref{correct MF density matrix}) into (\ref{entropy MF}). Thus, the constrained minimum of $\mathcal{F}(\vec{A}, \vec{\lambda})$ corresponds to the constrained maximum of $\mathcal{S}_{\lambda}( \hat{\rho}_{\lambda}(\vec{A}, \vec{\lambda}), \vec{A}, \vec{\lambda})$, here understood as a function of $\vec{A}$ and $ \vec{\lambda}$ only,  once Eqs. (\ref{derivative omega}) and (\ref{derivative q i}) are solved. From now on we will use $\mathcal{F}$ instead of $\mathcal{S}_{\lambda}$.
 With the help of (\ref{mathcal F functional}), the Eqs. (\ref{mean value of H GMF with lambda, d/dA}) and (\ref{mean value of H GMF with lambda, d/dlambda}) are equivalent to 
\begin{equation}
\nabla_{A} \mathcal{F} =0, ~~~~~ \nabla_{\lambda} \mathcal{F} =0, 
\label{derivative of mathcalF A, lambda}  
\end{equation}
where $\nabla_{A} \equiv \big( \frac{\partial }{\partial  A_{1}}, \frac{\partial }{\partial  A_{2}}, \ldots, \frac{\partial }{\partial  A_{M}}  \big)$, and analogously for $\nabla_{\lambda}$.
When deriving the last condition we made use of the identity (c.f. \cite{Zubariev})
\begin{equation}
\frac{\partial }{\partial x} e^{C(x)} = \int_{0}^{1} d\tau e^{  \tau C(x)} \Big (\frac{\partial }{\partial x} C(x) \Big)  e^{ (1- \tau) C(x)}
\label{general derivative of exp op},
\end{equation}
in combination with the linearity of the trace operation, as well as  its invariance with respect to cyclic permutation of the operators. 

For each solution of (\ref{derivative omega})-(\ref{derivative lambda s}) or, equivalently, of (\ref{derivative of mathcalF A, lambda}), it should be verified  separately  if  also the sufficient  conditions for the existence of the constrained  minimum of $\mathcal{F}$ are fulfilled. In practice, this will be most conveniently  carried out through numerical analysis.
 
We may expect that one of the solutions  of  (\ref{derivative of mathcalF A, lambda}), labeled  ($\vec{A}_0$,  $\vec{\lambda}_0$), yields such desired  global  minimum of $\mathcal{F}$   subject to constraints,  and this defines the equilibrium state. If  more than one solution with the minimal  value of  $\mathcal{F}$ exist, we select one of them, breaking the remaining  symmetry.

However,  $\mathcal{F}$ is well defined also for the values of $\vec{A}$ which differ from $\vec{A}_0$, and  may possess physical interpretation provided that the consistency conditions (\ref{mean value of H GMF with lambda, d/dlambda}) are met. Namely,  $\mathcal{F}(\vec{A}, \vec{\lambda}(\vec{A}))$ may be interpreted as a Landau functional corresponding to the given MF model.  The explicit analytical form of $\mathcal{F}(\vec{A}, \vec{\lambda}(\vec{A}))$ or  its series expansion in the powers of order parameter(s) may be difficult to obtain, and then again we have to resort to the  numerical analysis. Nonetheless,  the concepts of Landau theory of phase transitions may be utilized. E.g., the quantity $\mathcal{N}^{-1}\exp\big(-\beta  \mathcal{F}(\vec{A}, \vec{\lambda}(\vec{A}))\big)$, where $\mathcal{N}$ is normalization constant, may be interpreted as the probability distribution  of $\vec{A}$. Also, plots of $\mathcal{F}(\vec{A}, \vec{\lambda}(\vec{A}))$ may be used for qualitative analysis of the character of the emerging phase transition. % Such a point of view is not allowed within the standard MF approach, as then (a) few particular values/ isolated points is usually  of each of the mean-fields is selected by equations (\ref{mean value of A set of eq}). The more detailed analysis of this problem is carried out elsewhere \cite{Jedrak}.

\subsection{Mean-field equilibrium thermodynamics\label{Thermodynamics}}
%\footnote{Strictly speaking, the $\Lambda \to \infty$ limit corresponds implies also $D \to \infty$ limit, where $D$ is dimension of th Hilbert space. Although we assume, that $D$ is finite. Nonetheless, finite, but  large enough $D$ will/would be /is sufficient.}
The equilibrium description is justified, provided that at the point $\vec{A} = \vec{A}_0$ the function ~$\mathcal{N}^{-1}\exp\big(-\beta  \mathcal{F}(\vec{A}, \vec{\lambda}(\vec{A}))\big)$ possesses a sharp peak, and consequently, the fluctuations of order parameter can be neglected. This condition is expected to hold for a wide class of  MF Hamiltonians defined on lattice, in the $\Lambda \to \infty$ limit, where $\Lambda$ is the number of lattice sites. Obviously, for such models, this condition is equivalent to the thermodynamic limit $ V \to \infty$, $N \to \infty$ with finite $N/V = n$. Then we  can define the thermodynamic grand potential 
\begin{equation}
\Omega(T, V, \mu) = \mathcal{F}(T, V, \mu; \vec{A}_0(T, V, \mu), \vec{\lambda}_0(T, V, \mu)). 
\label{Omega thermo MF}
\end{equation}
With this definition, the MF thermodynamics can be constructed, but we still have to demonstrate its consistency with MF statistical mechanical description based on (\ref{correct MF density matrix}), cf. \cite{Argyres Kaplan}. Such consistency is certainly present for an exact description, making use of non-mean-field Hamiltonians \cite{statistical mechanics, Zubariev, Jaynes, Argyres Kaplan}; in that case we have for example
\begin{equation}
S = -\Big( \frac{\partial \Omega}{\partial T}\Big)_{V, \mu}, ~~~ p = -\Big( \frac{\partial \Omega}{\partial V}\Big)_{T, \mu} , ~~~ N = -\Big( \frac{\partial \Omega}{\partial \mu}\Big)_{T, V}. 
\label{S deriv Omega}
\end{equation}
One may ask if in the MF case the r.h.s. of (\ref{S deriv Omega}) correspond to their statistical-mechanical definitions, as the equilibrium MF Hamiltonian $\hat{H}_{\lambda}$  obviously depends on temperature and chemical potential through its dependence on  the equilibrium values of mean fields and Lagrange multipliers\footnote{  
We exclude the possibility of explicit dependence of $\hat{H}_{\lambda}$ on $\beta = 1/T$ or $\mu$.}, $(\vec{A}_{0},\vec{\lambda}_{0})$.
It can  be shown, that (\ref{S deriv Omega}) indeed holds, due to the underlying variational principle. Explicitly, %\beta^{2}  \Big( \frac{\partial \beta^{-1} \ln \mathcal{Z}_{\lambda}}{\partial \beta}\Big)_{V, \mu}
\begin{eqnarray}
\Big( \frac{\partial \Omega}{\partial T}\Big)_{V, \mu} &=&  \beta^{2} \partial_{ \beta}\big( \beta^{-1} \ln \mathcal{Z}_{\lambda}\big) = \beta \mathcal{Z}_{\lambda}^{-1}  \partial_{ \beta} \mathcal{Z}_{\lambda} - \ln \mathcal{Z}_{\lambda} \nonumber \\ 
&=& - \beta \mathcal{Z}_{\lambda}^{-1} \text{Tr}[ (\beta \partial_{ \beta} \hat{K}_{\lambda} + \hat{K}_{\lambda}) e^{-\beta \hat{K}_{\lambda}} ] - \ln \mathcal{Z}_{\lambda} \nonumber \\ 
&=& - \beta^{2}  \langle \partial_{ \beta} \hat{K}_{\lambda}\rangle_{\lambda} - \beta \langle \hat{K}_{\lambda} \rangle_{\lambda}  - \ln \mathcal{Z}_{\lambda} \nonumber \\ 
&=& - \beta^{2}  \langle \partial_{ \beta} \hat{K}_{\lambda}\rangle_{\lambda}  + \langle \ln \hat{\rho}_{\lambda}\rangle_{\lambda} = - \beta^{2}  \langle \partial_{ \beta} \hat{K}_{\lambda}\rangle_{\lambda} -S, 
\end{eqnarray}
where $\hat{K}_{\lambda} = \hat{H}_{\lambda} -\mu \hat{N}$ and $\mathcal{Z}_{\lambda}$ are evaluated for $\vec{A}= \vec{A}_{0}$, $\vec{\lambda}= \vec{\lambda}_{0}$. Also, $\ln \mathcal{Z}_{\lambda}(\vec{A}_{0},\vec{\lambda}_{0}) = -\beta \Omega $, and $\langle \hat{K}_{\lambda}\rangle_{\lambda} = U - \mu N$. In effect, we obtain the desired result (\ref{S deriv Omega}), provided that the term  $\langle \partial_{ \beta} \hat{K}_{\lambda}\rangle_{\lambda}$ vanishes. Indeed, in equilibrium we have
\begin{eqnarray}
\langle \partial_{ \beta} \hat{K}_{\lambda}\rangle_{\lambda}&=& \Big \langle \sum_{s = 1}^{M}\Big( \frac{\partial  \hat{K}_{\lambda}}{\partial A_s} \frac{\partial A_{s0}}{\partial \beta} + \frac{\partial  \hat{K}_{\lambda}}{\partial \lambda_s} \frac{\partial \lambda_{s0}}{\partial \beta}\Big)\Big \rangle_{\lambda} \nonumber \\ 
&=&  \sum_{s = 1}^{M}\Big( \Big \langle \frac{\partial  \hat{H}_{\lambda}}{\partial A_s}\Big \rangle_{\lambda} \frac{\partial A_{s0}}{\partial \beta} + \Big \langle \frac{\partial  \hat{H}_{\lambda}}{\partial \lambda_s} \Big \rangle_{\lambda} \frac{\partial \lambda_{s0}}{\partial \beta}\Big) = 0, \nonumber \\ 
\end{eqnarray}
due to relations (\ref{mean value of H GMF with lambda, d/dA}) and (\ref{mean value of H GMF with lambda, d/dlambda}). The same reasoning applies to the derivatives of $\Omega$ with respect to  $\mu$ and $V$. In the latter case, Hamiltonian depends explicitly on $V$ (e.g. through the volume dependence of its parameters). In effect, we recover proper expressions for average number of particles $N$ and pressure $p$. Other thermodynamic potentials may be constructed in the standard manner, e.g. $F = \Omega + \mu N$. 

Note that as the $\beta$- and $\mu$-  dependences of equilibrium values of mean-fields $\vec{A}_{0}$ obtained through self-consistent Eqs. (\ref{mean value of A set of eq}) are nontrivial in general, and also because (\ref{mean value of H GMF with lambda, d/dA}) with $\vec{\lambda} = \vec{0}$ usually does not  hold in such case, there is no reason for $\langle \partial_{ \beta} \hat{K}\rangle$ (as well as for analogous expressions for the derivatives with respect to $\mu$, $V$ etc.) to vanish. This feature results in the lack of consistency between MF thermodynamics and MF statistical mechanics even on the level of the first-order derivatives of the corresponding thermodynamic potential and constitutes an obvious and serious drawback of  non-variational self-consistent approach\footnote{A notable exception, the case of pure Hartree or Hartree-Fock type of MF Hamiltonians, is discussed in the next paragraph.}. 
The problem of such consistency is also analyzed in Ref. \cite{Argyres Kaplan}, however, from a slightly different perspective. 
The analysis of the second-order derivatives of  $\Omega$ is slightly more complicated and is carried out in \cite{Jedrak}.

%At this stage,  we return to the problem of solving Eqs. (\ref{derivative of mathcalF A, lambda}).  In practice, the most convenient way would  probably be to work in the eigenbasis of $\hat{\rho}_{\lambda}$, as then the explicit form of $\mathcal{F}$ as a function of $\vec{A}$ and $\vec{\lambda}$ can be obtained. Obviously, Eqs. (\ref{derivative of mathcalF A, lambda}) may have many solutions, each of them representing different physical state, which  depends on values of $\beta,\mu$ and  parameters appearing in the Hamiltonian. In such situation the right choice is obviously the one minimizing $\mathcal{F}$.

%Let us note, that the formalism presented here may also be applied to the spin systems, with the only difference, that in that situation the canonical rather than the grand-canonical ensemble should be used instead. In particular, this formalism can be used to build up a mean-field description for the case of the classical spin and spin-like models.  In such a situation we deal with a probability distribution, which may obviously be represented by the density matrix, that can be diagonalized simultaneously with all observables appearing in the problem at hand. 
%The presence of source  terms $ \tilde{\lambda}_{s}  \big( (\hat{A}_{s})_{ii} -  A_{s})q_{i} \big)$ in $\hat{H}_{\lambda}$ allows one to compute, at least for  mean field correlations  

%%%%%%%%%%%% R E D U N D A N C Y

\subsection{Redundancy conditions  \label{Hartree}}
The natural question is, when the presence of Lagrange multipliers in the Hamiltonian (\ref{GMF Hamiltonian with lambdas}) is unnecessary. We will not provide the general answer, but instead present an important example. Suppose we have the exact  Hamiltonian  of the following form
\begin{equation}
\hat{H}_{e} = \hat{H}_{0e} + \sum_{\{\kappa, \gamma \}} V_{\kappa \gamma} \hat{\mathcal{O}}_{\kappa}\hat{\mathcal{O}}_{\gamma},  
\label{no lambda H parent}
\end{equation}
where each  $\hat{\mathcal{O}}_{\kappa}$ is assumed to be bilinear in the creation and/or   
 annihilation operators, and the summation is taken over  all  pairs $\{\kappa, \gamma\}$ of multi indices.
If the interaction term in  (\ref{no lambda H parent}) is decoupled  (i.e. replaced by its MF counterpart)  according to\footnote{Usually, the $\hat{\mathcal{O}}_{\kappa}\hat{\mathcal{O}}_{\gamma}$ term may be decoupled in more then one way, but this does not change the arguments presented.} 
\begin{equation}
 \hat{\mathcal{O}}_{\kappa}\hat{\mathcal{O}}_{\gamma} \to     \hat{A}_{s} A_{t}  + A_{s} \hat{A}_{t} - A_{s} A_{t}, 
\label{AB BF}
\end{equation} 
with $A_s$, $A_t$ given by (\ref{mean value of A set of eq}) for the appropriate density matrix, then the standard Hartree or the Hartree-Fock (HF) type mean-field Hamiltonian is obtained. With  the constraint terms it reads
\begin{eqnarray}
\hat{H}_{\lambda} &=&  \hat{H}_{0e} +  \sum_{\{s,t\}} \tilde{V}_{st} \big( \hat{A}_{s} A_{t}  + A_{s} \hat{A}_{t} - A_{s} A_{t}\big)  - \sum_{s}  \lambda_s (\hat{A}_{s}  - A_{s}) \nonumber \\
&\equiv&  \hat{H}  - \sum_{s}  \lambda_s (\hat{A}_{s}  - A_{s}).
\label{no lambda H}
\end{eqnarray}
In the above,  $\hat{H}_{0e}$ does not depend on $\vec{A}$, and  $\tilde{V}_{st}$ are the  combinations of matrix elements $V_{\kappa \gamma}$. Then, one can find that 
\begin{equation}
 \Big \langle  \frac{\partial  \hat{H} }{\partial A_t} \Big \rangle_{\lambda}  =  \Big \langle \sum_{s}\tilde{V}_{st} \big( \hat{A}_{s} - A_{s}\big)  \Big \rangle_{\lambda}=  0,  
\label{mean value of H GMF with lambda, d/dA HF type}
\end{equation}
which together with (\ref{mean value of H GMF with lambda, d/dA}) implies that $\lambda_t =0$, for any $t$. Also, for  Hamiltonian of such form, the solution of   $M$ self-consistent equations (\ref{mean value of A set of eq}) coincide with those of  $M$ equations $\nabla_{A} \mathcal{F} =0 $ of unconstrained variational approach. A related discussion of this problem  may be found in \cite{FB}.

Summarizing, for any MF Hamiltonian of the Hartree-Fock type,  application of our method is not necessary. %The same reasoning applies to the situation when only the Hamiltonian has the form (\ref{no lambda H}) with respect to  some of $\{ A_s \}$ variables. 
However, there are important examples of the MF Hamiltonians not being of HF form, in Section \ref{examples} we present two such  examples.
 
\subsection{Relation of the present method to the variational principle of Bogoliubov and Feynman \label{Bogoliubov}}

In the previous Sections we considered  the situation, when the description of the system has been based  \textit{solely} on  the MF Hamiltonian $\hat{H}$. In other words, we made no use of any exact Hamiltonian  $\hat{H}_{e}$ present in a problem at hand,  from which  $\hat{H}$ may be derived and for  which it  may play the role of  a simplified counterpart. %Such a treatment may seem questionable because one can argue, that without the exact  Hamiltonian being used to build a density operator, we have no real basis for the analysis.%, in particular, resulting values of  energy (or other thermodynamic potentials) may be lower  then what we expect for the exact Hamiltonian. %On the other hand, resulting MF Hamiltonian may be not much more unrealistic than $\hat{H}_{e}$, which by itself is  simplified and  covers  only a part of the real-system complexity.

However, frequently the MF density operators are used  as a trial variational states in order  to determine the bounds on the grand potential $\Omega_e$ or free energy $F_e$ of the system described by  $\hat{H}_{e}$. Such bounds are provided by the well-known inequality due to Bogoliubov and Feynman \cite{Feynman, Decoster} 
\begin{equation}
\Omega_{e} \leq \langle\hat{K}_{e} - \hat{K}_0 \rangle_0 +  \tilde{\Omega} = \text{Tr}[ \hat{\rho}_0 (\hat{K}_{e} - \hat{K}_0)] +   \tilde{\Omega}. 
\label{Bogoliubov GC}
\end{equation}
In the above, $\hat{H}_0$ is a trial Hamiltonian, $\tilde{\Omega} = -\beta^{-1}\ln \text{Tr}[ \exp \big( - \beta( \hat{H}_0-\mu_0 \hat{N}) \big)]$, and $\Omega_e = -\beta^{-1}\ln \text{Tr}[ \exp \big( - \beta( \hat{H}_e-\mu_e \hat{N}) \big)]$. Also, $\hat{K}_{e}=\hat{H}_e-\mu_e \hat{N}$ and analogously for $\hat{K}_0$, with $\mu_e$ ($\mu_0$) being respective chemical potentials. There are several ways to prove (\ref{Bogoliubov GC}), probably the shortest of them  is that of making use of the Klein inequality, \cite{Nielsen} 
\begin{equation}
\text{Tr}[ \hat{\rho} \ln  \hat{\rho}] \geq \text{Tr}[ \hat{\rho} \ln  \hat{\sigma}], 
\label{Klein e}
\end{equation}
which holds for any normalized density operators $\hat{\rho}$ and $ \hat{\sigma}$. Inserting 
\begin{equation}
\hat{\rho} = \hat{\rho}_0 = Z_0^{-1} \exp(-\beta(\hat{H}_0 -\mu_0 \hat{N})),  ~~~~~ \hat{\sigma} =\hat{\rho}_{e} = Z_{e}^{-1} \exp(-\beta(\hat{H}_{e} -\mu_e \hat{N})),
\label{}
\end{equation}
(with $Z_0 = \text{Tr}[\exp(-\beta(\hat{H}_0 -\mu_0 \hat{N}))]$ and $Z_e = \text{Tr}[\exp(-\beta(\hat{H}_e -\mu \hat{N}))]$) into (\ref{Klein e}),  we obtain (\ref{Bogoliubov GC}).  %$\Omega = \mathcal{F}$
If one wants to work with the chemical potential as an independent variable, then $\mu_e = \mu_0 \equiv \mu$ and  
(\ref{Bogoliubov GC}) reduces to 
\begin{equation}
\Omega_{e} \leq \langle\hat{H}_{e} - \hat{H}_0 \rangle_0 +  \tilde{\Omega} = \text{Tr}[ \hat{\rho}_0 (\hat{H}_{e} - \hat{H}_0)] +   \tilde{\Omega}. 
\label{Bogoliubov GC mu}
\end{equation}
However, usually it is more convenient to have particle number $N$ as an independent variable. In such a case $ \mu_0$  and (unknown) $ \mu_e$ are not equal in general, as they are determined from different conditions, i.e. $\langle \hat{N}\rangle_0 = N $ and $\langle \hat{N}\rangle_e = N $, respectively. However, (\ref{Bogoliubov GC}) may be then given the form
\begin{equation}
F_e = \Omega_{e} + \mu_e N   \leq \langle\hat{H}_{e} - \hat{H}_0 \rangle_0 +   \tilde{\Omega}+\mu_0 N. 
\label{Bogoliubov GC N}
\end{equation}
% \tilde{\Omega}+\mu_0 N =  \tilde{F}
Bogoliubov inequality is the basis of the variational principle, the best  trial state  $\hat{\rho}_0$  is the one  minimizing r.h.s.  of (\ref{Bogoliubov GC mu}), or (\ref{Bogoliubov GC N}), depending on the choice of independent variables, $\mu$ or  $N$, respectively. 

If one wants to use the Bogoliubov inequality in order to determine upper bounds for $\Omega_e$ or $F_e$ of some  $\hat{H}_{e}$, but with the   self-consistency of the approach being preserved, then the density operators (\ref{MF naive DO}) are not good candidates for the trial states. Clearly, they have no free variational parameters\footnote{We neglect the situation, in which there appear some extra variational parameters, not being expectation values of operators present in a problem, and thus not requiring the corresponding Lagrange multipliers.}
 if only the values of mean-fields, $\vec{A}=\vec{A}_0$, are obtained in a self-consistent fashion through (\ref{mean value of A set of eq}). On the other hand, if  $\vec{A}_0$  were obtained from unwary minimization of the r.h.s. of (\ref{Bogoliubov GC mu}), (\ref{Bogoliubov GC N})  with  $\hat{\rho}$ given by (\ref{MF naive DO}), then  the self-consistency of the approach would  be violated in general (again with the notable exception of the Hartree-Fock MF Hamiltonians).

In contrast, consider the trial state chosen in the form (\ref{correct MF density matrix}), with the trial Hamiltonian  $\hat{H}_{0} = \hat{H}_{\lambda}$, yet unrelated to the $\hat{H}_{e}$ in question. Then the r.h.s. of (\ref{Bogoliubov GC mu}) reads 
\begin{equation}
\text{Tr}[ \hat{\rho}_{\lambda} (\hat{H}_{e} - \hat{H}_{\lambda})] + \mathcal{F}(\vec{A}, \vec{\lambda}) \equiv  \mathcal{B} (\vec{A}, \vec{\lambda}),
\label{Bogoliubov GC for rho lambda mu}
\end{equation}
whereas that of (\ref{Bogoliubov GC N}), respectively (with $\mu_0 \equiv \mu$)
\begin{equation}
\text{Tr}[ \hat{\rho}_{\lambda} (\hat{H}_{e} - \hat{H}_{\lambda})] + \mathcal{F}(\vec{A}, \vec{\lambda}) + \mu N =  \mathcal{B} (\vec{A}, \vec{\lambda}) + \mu N \equiv \mathcal{B}_N (\vec{A}, \vec{\lambda}).
\label{Bogoliubov GC for rho lambda N}
\end{equation}
Depending on the choice of independent variable, the necessary\footnote{Similarly as previously, here we assume, that the desired minimum corresponds to  a stationary point.} $2M$ conditions for r.h.s of the Bogoliubov inequality, i.e $\mathcal{B}$ or $\mathcal{B}_N $, to have a  minimum subject to constrains of self-consistency acquire slightly different form. For (\ref{Bogoliubov GC for rho lambda mu}) we have
\begin{equation}
\nabla_{A} \mathcal{B}(\vec{A}, \vec{\lambda}) =0, ~~~~~~~   \nabla_{\lambda} \mathcal{F}(\vec{A}, \vec{\lambda})=0.
\label{Bogoliubov set of eq}
\end{equation}
On the other hand, for (\ref{Bogoliubov GC for rho lambda N}) we have 
\begin{equation}
\nabla^{\prime}_{A} \mathcal{B}_N(\vec{A}, \vec{\lambda}) =0, ~~~~~~~   \nabla_{\lambda}(\mathcal{F}(\vec{A}, \vec{\lambda}) + \mu N)=0, ~~~~ \frac{\partial}{\partial N}(\mathcal{F}(\vec{A}, \vec{\lambda}) + \mu N)= \mu.
\label{Bogoliubov set of eq N}
\end{equation}
In the above, $\nabla^{\prime}_{A}$ labels the gradient taken with respect to all mean-fields except $A_1 = N$. Also  
\begin{equation}
\nabla^{\prime}_{A}\mathcal{B}_N(\vec{A}, \vec{\lambda})=\nabla^{\prime}_{A}\mathcal{B}(\vec{A}, \vec{\lambda}), ~~~ ~~ \nabla_{\lambda}(\mathcal{F}(\vec{A}, \vec{\lambda}) + \mu N)=\nabla_{\lambda}\mathcal{F}(\vec{A}, \vec{\lambda}).
\label{prime is not a prime}
\end{equation}

If $\mathcal{B}(\vec{A}, \vec{\lambda}) \neq \mathcal{F}(\vec{A}, \vec{\lambda})$, the solution of (\ref{Bogoliubov set of eq}) or (\ref{Bogoliubov set of eq N}) leads  to different values of $\vec{A}_{0}$ and $\vec{\lambda}_{0}$, than those obtained from (\ref{derivative of mathcalF A, lambda}). Then,  $\hat{\rho}_{\lambda}$ is   no longer a true grand-canonical MF density operator for $\hat{H}_{\lambda}$. The above  remarks hold  for other inequalities (i.e. various generalizations of (\ref{Bogoliubov GC})), and the corresponding variational principles built on them, \cite{Decoster}. One may use  the $\hat{\rho}_{\lambda}$  (\ref{correct MF density matrix}) as useful ans\"{a}tze in such cases as well. Nonetheless,  the entropy functional (\ref{entropy MF}) with the corresponding variational principle plays a special role, and it  determines the form (\ref{correct MF density matrix})  of GC MF density operator. Also, the self-consistency conditions may be conveniently formulated with the help of $\mathcal{F}(\vec{A}, \vec{\lambda})$, as in (\ref{Bogoliubov set of eq}) and (\ref{Bogoliubov set of eq N}).
However, clearly, for any  pair of $\hat{H}_{e}, \hat{H}_{\lambda}$, for which the condition  
\begin{equation}
\langle \hat{H}_{e}-\hat{H}_{\lambda} \rangle_{\lambda} =0
\label{}
\end{equation}
 holds, so consequently also $\tilde{\Omega}(\vec{A}, \vec{\lambda}) = \mathcal{F}(\vec{A}, \vec{\lambda}) = \mathcal{B}(\vec{A}, \vec{\lambda})$, the Bogoliubov - Feynman and Maximum Entropy  variational principles coincide. In such case the optimal, from the point of view of maximum entropy inference, i.e. the most uninformative MF density operator $\hat{\rho}_{\lambda}$ provides us also with the upper bound for exact grand potential $\Omega_e$  or free energy $F_e$ of $\hat{H}_{e}$. For (\ref{Bogoliubov set of eq}) this is obvious. For (\ref{Bogoliubov set of eq N}), this may be shown using $\partial\mathcal{F}/ \partial\mu = -N$  and replacing $\mathcal{B} \to \mathcal{F}$, then we obtain the identity 
\begin{eqnarray}
\mu &= & \frac{\partial}{\partial N}(\mathcal{F}(\vec{A}, \vec{\lambda}) + \mu N)= \frac{\partial\mathcal{F}}{\partial N} + \frac{\partial\mathcal{F}}{\partial \mu} \frac{\partial\mu}{\partial N} + N \frac{\partial\mu}{\partial N} = \frac{\partial\mathcal{F}}{\partial N} + \mu.
\label{prove F}
\end{eqnarray}
Consequently, $\partial\mathcal{F}/ \partial N=0$ for the solutions of Eqs. (\ref{Bogoliubov set of eq N}).
From Eqs. (\ref{prime is not a prime}) and (\ref{prove F}), it follows that any solution of  (\ref{Bogoliubov set of eq N}), i.e. $(\vec{A}_0, \vec{\lambda}_0, \mu)$ is  also a solution of (\ref{Bogoliubov set of eq}), if  $N$ and $\mu$ related by the condition $\langle \hat{N} \rangle = N = -\partial \mathcal{F}/\partial \mu$. Also, conversly, Eqs. (\ref{prime is not a prime}), (\ref{prove F}) imply that  the solutions of (\ref{Bogoliubov set of eq}) are also solutions of  (\ref{Bogoliubov set of eq N}),  thus both schemes  are equivalent. 
%%%%%%%%%%%%%%%%%%%% E X A M P L E S %%%%%%%%%%%%%%%%%%%%%%%%%%%%%%%%%%%%%%%% \footnote{Note, that $\langle \hat{H} \rangle_{\lambda} = \langle \hat{H}_{\lambda} \rangle_{\lambda} $ if the self-consistency conditions are met.}.
\section{Two nontrivial examples\label{examples}}
We now discuss briefly two MF models, for which the proposed method leads to  nontrivial corrections. To show this, we make use of Eq. (\ref{mean value of H GMF with lambda, d/dA}), and  argue, that the condition $\vec{\lambda}=\vec{0}$ cannot be, in general, satisfied.  %More detailed analysis  will be carried out elsewhere \cite{Jedrak Spalek}.  

\subsection{Renormalized  t-J model \label{tJ}}

The first example we analyze is the MF Hamiltonian  of the so-called renormalized t-J model 
\cite{ZGRS}-\cite{LiZhouWang}. It originates from the standard t-J model \cite{Spalek Oles},  believed to describe correctly the essential physics of the cuprate high-temperature superconductors, that is expressed by the Hamiltonian
\begin{equation}
 \hat{H}_{e} = \sum_{i, j(i), \sigma} t_{ij} \tilde{c}_{i \sigma}^{\dag} \tilde{c}_{j \sigma} + J \sum_{\langle i j \rangle}  ( \mathbf{S}_{i}\cdot \mathbf{S}_{j} - \frac{1}{4}\hat{\nu}_i \hat{\nu}_{j}) -\mu \sum_{i,  \sigma} \hat{\nu}_{i \sigma}.
\label{t-J exact}
\end{equation}
Here $\tilde{c}_{i \sigma}= (1- n_{i -\sigma} )c_{i \sigma}$ and $ \hat{\nu}_i = \sum_{\sigma}= \tilde{c}^{\dag}_{i \sigma}\tilde{c}_{i \sigma}$ are the operators with double occupancies on site $i$ projected out. The corresponding renormalized MF Hamiltonian may be taken in the form \cite{Didier, Marcin 2}
\begin{eqnarray}
\hat{H} &=&  \sum_{\langle i j \rangle \sigma}\Big( \big (t_{ij} g^{t}_{ij} c_{i \sigma}^{\dag} c_{j \sigma} +  \text{H.c.} \big)  - \frac{3}{4} J g^{J}_{ij} (\chi_{ji}  c_{i \sigma}^{\dag} c_{j \sigma} + \text{H.c.} - |\chi_{ij}|^{2}) \nonumber \\ &+& - \frac{3}{4}J g^{J}_{ij} (\Delta_{ij}   c^{\dag}_{j \sigma}c_{i -\sigma}^{\dag} + \text{H.c.}   - |\Delta_{ij}|^{2})\Big)  -\mu  \sum_{i \sigma}   c_{i \sigma}^{\dag} c_{i \sigma},
\label{ren tJ Ham}
\end{eqnarray}
with  $n_i = \sum_{\sigma} \langle  c_{i \sigma}^{\dag}  c_{i \sigma} \rangle $,   $\chi_{ij} = \langle  c_{i \sigma}^{\dag}  c_{j \sigma} \rangle$, and $\Delta_{ij} = \langle  c_{i -\sigma}  c_{j \sigma} \rangle = \langle  c_{j -\sigma}  c_{i \sigma} \rangle$. 
In this form the projections have been abolished at the price of introducing the expectation-value dependent renormalization factors $g^{t}_{ij} $ and $g^{J}_{ij} $ resulting from the Gutzwiller approximation \cite{Vollhardt}. %Their various detailed forms will be given/presented below. %We assume also $\Delta_{ij} = \Delta^{\ast}_{ij}$.  
%The attempt for/ to construct a GC density operator for (\ref{ren tJ Ham}) may seem questionable, as the renormalization factors were devised for $T=0$ case/situation. However, it is frequently used also in the/ to study/discuss/examine the  situation with $T\neq 0$,  which is also studied below, as for low temperatures its application seems to be justified. 
At first, we consider  the renormalization factors  depending solely on  $x_i = 1- n_i$, i.e.
\begin{equation}
 g^{t}_{ij} = \sqrt{\frac{4x_i x_j}{(x_i + 1)(x_j + 1)}} ~~~~\text{and} ~~~ g^{J}_{ij} = \frac{4}{(x_i + 1)(x_j + 1)}.
\label{g ren fac x only}
\end{equation} 
The task is  to  determine the values of the mean-field parameters
$\chi_{ij}$, $\Delta_{ij}$ and $x_{i}$. This could be  achieved with the help of self-consistent Eqs. (\ref{mean value of A set of eq}) for $\hat{H}$ given by (\ref{ren tJ Ham}) (cf. \cite{ZGRS}- \cite{Marcin 2}).%, without redefinition of the Hamiltonian given by/according to (\ref{GMF Hamiltonian with lambdas}). 

On the other hand, we may proceed in the framework of  our method, and redefine the Hamiltonian (\ref{ren tJ Ham}) according to
\begin{eqnarray} % \langle i j \rangle
\hat{H}_{\lambda} = \hat{H} &-& \sum_{i} \lambda^{n}_{i}( \sum_{\sigma} c_{i \sigma}^{\dag}  c_{i \sigma} - n_{i}) - \sum_{\langle i j \rangle \sigma}  \big( \lambda^{\chi}_{ij}  (c_{i \sigma}^{\dag}  c_{j \sigma} - \chi_{ij}) + \text{H.c.} \big) \nonumber \\ &- &  \sum_{\langle i j \rangle \sigma} \big( \lambda^{\Delta}_{ij}  (c_{i - \sigma}  c_{j \sigma} - \Delta_{ij})  + \text{H.c.} \big).
\label{MF tJ lambda term}
\end{eqnarray}
Assuming no additional symmetries, using $\partial / \partial n_i = - \partial / \partial x_i$,  converting the sums over bonds to that over sites in  (\ref{ren tJ Ham}) according to  $2 \sum_{\langle i, j \rangle} = \sum_{i, j(i)} $ and finally taking $A_w \equiv n_i $ in Eqs.  (\ref{mean value of H GMF with lambda, d/dA}) we obtain 
%\begin{equation} \sum_{j} \big( 2 t_{ij} \frac{\partial g^{t}_{ij}}{\partial x_i} \text{Re}\chi_{ij} - \frac{3}{4}J \sum_{i j \sigma} \frac{\partial g^{J}_{ij}}{\partial x_i}  |\chi_{ij}|^{2})  = \lambda^{x}_{i}. \label{lambda x i zero} \end{equation}

\begin{eqnarray}
2 \lambda^{n}_{i} &=& \Big \langle \sum_{j(i), \sigma} t_{ij}  \frac{\partial g^{t}_{ij}}{\partial x_i} \big( c_{i \sigma}^{\dag} c_{j \sigma} + \text{H.c.} \big) \Big \rangle_{\lambda} \nonumber \\
&- & 
\Big \langle\frac{3}{4}J     \sum_{j(i), \sigma} \frac{\partial g^{J}_{ij}}{\partial x_i} (\chi_{ji}  c_{i \sigma}^{\dag} c_{j \sigma} + \text{H.c.} - |\chi_{ij}|^{2})\Big \rangle_{\lambda} \nonumber \\
&- &
\Big \langle   \frac{3}{4}J     \sum_{j(i), \sigma} \frac{\partial g^{J}_{ij}}{\partial x_i} (\Delta_{ij}   c^{\dag}_{j \sigma}c_{i -\sigma}^{\dag} + \text{H.c.}   - |\Delta_{ij}|^{2})\Big \rangle_{\lambda} \nonumber \\
&=& \sum_{j(i)} \Big( 4 t_{ij} \frac{\partial g^{t}_{ij}}{\partial x_i} \text{Re}\chi_{ij} - \frac{3}{2}J  \frac{\partial g^{J}_{ij}}{\partial x_i} \big( |\chi_{ij}|^{2} + |\Delta_{ij}|^{2}  \big) \Big).
\label{lambda x i zero}
\end{eqnarray}
Below it will be shown that in this particular case, $\lambda^{\chi}_{ij}  = \lambda^{\Delta}_{ij}= 0$, for each bond $\langle ij \rangle$.
Consequently, if  for all $i$, we put $ \lambda^{n}_{i} = 0$, then the density operator, and all averages coincide with those of the standard self-consistent MF treatment. However, in such situation  equations (\ref{lambda x i zero}) and  the self-consistent equations (\ref{mean value of A set of eq}) for $\hat{H}$ given by (\ref{ren tJ Ham}) cannot  in general  be simultaneous satisfied. For example, if one sets   $J=0$ and $  x_{i}\in (0, 1)$, then from  (\ref{lambda x i zero}) it follows that%( $\frac{\partial }{\partial x_i} g^{t}_{ij} \neq 0$)     
\begin{equation} 
\Big \langle  \sum_{j(i) \sigma} \big( \frac{\partial g^{t}_{ij}}{\partial x_i}  c_{i \sigma}^{\dag} c_{j \sigma} +  \text{H.c.} \big)  \Big  \rangle  = 0.
\label{}
\end{equation}
This is a senseless result, as in such case our Hamiltonian reduces essentially to that of free fermions, at least  for  the  case of homogeneous solution, $x_i \equiv x$. Similarly, for $t_{ij} = 0$, one obtains  $ |\chi_{ij}|^{2} = 0$, $ |\Delta_{ij}|^{2} = 0$ for each bond $ij$, provided that each $t_{ij}$ is of the same sign,  as  $\partial g^{t}_{ij}/\partial x_i > 0$, for any $x_i$ and $\langle ij \rangle$. This is in contradiction  with the numerical results of Raczkowski \textit{et al.}, \cite{Marcin 1}. In other words, in this particular case, we can infer about the predictions of our method using the previous, non-variational  results. Consequently, it follows that $\lambda^{n}_{i}\neq 0$, for every $i$. 

From the point of view of our findings, a related situation can be found in Ref. \cite{LiZhouWang}. Namely, the inhomogeneous solutions of the model (\ref{ren tJ Ham}) (with slightly different definitions of  $\chi_{ij}$ and $\Delta_{ij}$), and with the renormalization factors given by (\ref{g ren fac x only}) are analyzed. However, the additional terms of the form $\sum_{i}\epsilon_i (\sum_{\sigma}  c_{i \sigma}^{\dag} c_{i \sigma} + x_i - 1)$, are included, where $\epsilon_i$ is the local fugacity. The quantity $\epsilon_i$  plays a role analogous to $ \lambda^{n}_{i}$ in our method. Inclusion of that additional constraint allows to treat each  ${x}_{i}$ as a variational parameter, and leads to the equations analogous to (\ref{lambda x i zero}). This  should be clear in the light of the discussion in the preceding Sections. 

Next, we pose the question of redundancy of the Lagrange multipliers $\lambda^{\chi}_{ij}$ and $\lambda^{\Delta}_{ij}$. We may expect, according to discussion carried out in Section \ref{Hartree},  that these quantities  vanish, as the Hamiltonian (\ref{ren tJ Ham}) with  $g^{t}_{ij}$ and $ g^{J}_{ij}$  given by (\ref{g ren fac x only}) is of the Hartree-Fock type with respect to both $\chi_{ij}$ and $\Delta_{ij}$. This is indeed the case, we may use once more  Eq. (\ref{mean value of H GMF with lambda, d/dA}) with $A_w = \chi_{ij} $ or $A_w = \Delta_{ij} $, respectively, to obtain
\begin{equation}
 \lambda^{\chi}_{ij} = \Big \langle   \frac{3}{4}J \sum_{\sigma}g^{J}_{ij} ( c_{j \sigma}^{\dag} c_{i \sigma} - \chi_{ij}^{\ast})\Big \rangle_{\lambda} \equiv 0, 
\label{HF like for chi}
\end{equation}

\begin{equation}
 \lambda^{\Delta}_{ij} = \Big \langle  \frac{3}{4}J \sum_{ \sigma}g^{J}_{ij} ( c_{j \sigma}^{\dag} c_{i -\sigma}^{\dag}  - \Delta_{ij}^{\ast})\Big \rangle_{\lambda}  \equiv 0.
\label{HF like for Delta}
\end{equation}
Obviously, those self-consistency equations must be fulfilled both in the previous treatments ($\vec{\lambda} = \vec{0}$) as well as in the present method.

Next, we consider  the renormalization factors of the form \cite{Marcin 2},
\begin{equation}
 g^{t}_{ij} = \sqrt{\frac{4x_i x_j(1-x_i)(1-x_j)}{(1-x_i^{2})(1-x_j^{2}) + 8(1-x_i x_j)|\chi_{ij}|^{2} + 16|\chi_{ij}|^{4}}},
\label{gt ren fac x chi delta}
\end{equation}
\begin{equation}
 g^{J}_{ij} = \frac{4(1-x_i)(1-x_j)  }{(1-x_i^{2})(1-x_j^{2}) + 8x_i x_j(|\Delta_{ij}|^{2}-|\chi_{ij}|^{2}) + 16(|\Delta_{ij}|^{4}+|\chi_{ij}|^{4})}.
\label{gJ ren fac x chi delta}
\end{equation}
 In such more general situation, the equation (\ref{lambda x i zero}) is still valid, and Eqs. (\ref{HF like for chi}) and  (\ref{HF like for Delta}) acquire  the form, respectively   

%%%%%%
\begin{eqnarray}
\lambda^{\chi}_{ij} &=& - \sum_{ \sigma} \Big \langle   t_{ij}  \frac{\partial g^{t}_{ij}}{\partial \chi_{ij}} \big( c_{i \sigma}^{\dag} c_{j \sigma} + \text{H.c.} \big)  -  \frac{3}{4}J  \frac{\partial g^{J}_{ij}}{\partial \chi_{ij}}  (\chi_{ji}  c_{i \sigma}^{\dag} c_{j \sigma} + \text{H.c.} - |\chi_{ij}|^{2})  \nonumber \\ &+& - \frac{3}{4}J g^{J}_{ij} ( c_{j \sigma}^{\dag} c_{i \sigma} - \chi_{ij}^{\ast}) - \frac{3}{4}J  \frac{\partial g^{J}_{ij}}{\partial \chi_{ij}}(\Delta_{ij} c^{\dag}_{j \sigma} c_{i -\sigma}^{\dag}  + \text{H.c.}  - |\Delta_{ij}|^{2}) \Big \rangle_{\lambda} \nonumber \\
&=& -  4 t_{ij} \frac{\partial g^{t}_{ij}}{\partial  \chi_{ij}} \text{Re}\chi_{ij} + \frac{3}{2}J  \frac{\partial g^{J}_{ij}}{\partial  \chi_{ij}} ( |\Delta_{ij}|^{2} + |\chi_{ij}|^{2}),
\label{lambda chi ij zero most general}
\end{eqnarray}
%%%%%%

\begin{eqnarray}
\lambda^{\Delta}_{ij} &=&  \sum_{\sigma} \Big   \langle   \frac{3}{4}J  \frac{\partial g^{J}_{ij}}{\partial \Delta_{ij}}  (\chi_{ji}  c_{i \sigma}^{\dag} c_{j \sigma} + \text{H.c.} - |\chi_{ij}|^{2}) +   \frac{3}{4}J g^{J}_{ij} (  c_{i -\sigma}^{\dag} c^{\dag}_{j \sigma} - \Delta_{ij}^{\ast}) \nonumber \\
&+&   \frac{3}{4}J  \frac{\partial g^{J}_{ij}}{\partial \Delta_{ij}} (\Delta_{ij} c^{\dag}_{j \sigma} c_{i -\sigma}^{\dag}  + \text{H.c.}  - |\Delta_{ij}|^{2})    \Big \rangle_{\lambda} \nonumber \\ &=& \frac{3}{2}J  \frac{\partial g^{J}_{ij}}{\partial  \Delta_{ij}} \big( |\Delta_{ij}|^{2} + |\chi_{ij}|^{2}\big). 
\label{lambda Delta ij zero most general}
\end{eqnarray} % 
Again, if we put $\vec{\lambda}=\vec{0}$ in order to reduce our approach  to the standard treatment, that Eqs. (\ref{lambda chi ij zero most general}) and (\ref{lambda Delta ij zero most general})
cannot be satisfied. For Eq. (\ref{lambda chi ij zero most general}) this can be  easily shown for the case  $t_{ij}=-t$ by adding it to its complex conjugate part. Then, at least for the class of solutions with  for $\text{Re}\chi_{ij}> 0$, one can convince oneself by explicitly computing respective derivatives of $g^{t}_{ij}$ and $ g^{J}_{ij}$, that the resulting expression should have nonzero value, for a wide range of values $x_i$. For Eq. (\ref{lambda Delta ij zero most general})  or (\ref{lambda chi ij zero most general}) in the case of $t_{ij}=0$,  no additional explanation is required. %computing explicitly respective derivatives of $g^{t}_{ij}, g^{J}_{ij}$.

Similar arguments apply if, instead of $\vec{\lambda}=\vec{0}$, a weaker conditions, $\lambda^{\chi}_{ij}=0$ or $\lambda^{\Delta}_{ij}=0$, are imposed. In such situation we can no longer make use of the numerical or analytical results  of the purely self-consistent approach. Nonetheless, it is unlikely that such conditions do not lead to contradictions. E.g. from  Eqs. (\ref{lambda chi ij zero most general}) with $t_{ij}=0$ or from  (\ref{lambda Delta ij zero most general}) in a general case,  we infer, that $ \chi_{ij} = \Delta_{ij}=0$.

In that slightly hand-waving manner we tried to argue, that for the Hamiltonian (\ref{ren tJ Ham}) our variational approach does not  reduce to that based solely on the self-consistency equations.

This model has been recently analyzed  and solved numerically in a separate publication \cite{Jedrak Spalek}.
 
%The interesting question then arises, which method provides us with a lower value of the free energy functional (\ref{mathcal F functional})? The answer to this question requires a detailed numerical investigation and will be given elsewhere \cite{Jedrak Spalek}. However, this point is not crucial for this discussion to hold.
%%%%%%%%%%%%%%%%
%%%%%%%%%%%%%%%%%%% S S M

\subsection{Electrons with spin-dependent effective masses}
In Ref.  \cite{Spalek Gopalan}, an approximate one-particle description of a strongly-correlated electron system was introduced, based on the observation,  that the effective masses of quasiparticles may be spin-dependent. The MF Hamiltonian of this model depends on $ A_{1} = N$ and $A_2 = M$, where 
\begin{equation}
 \hat{N} =  \hat{A}_1 = \sum_{\mathbf{k}} \sum_{\sigma} c^{\dag}_{\mathbf{k} \sigma}c_{\mathbf{k} \sigma}, ~~~~~  \hat{M} = \hat{A}_2 =\sum_{\mathbf{k}} \sum_{\sigma} \sigma c^{\dag}_{\mathbf{k} \sigma}c_{\mathbf{k} \sigma}.
\label{}
\end{equation}
With additional constraint terms, the Hamiltonian reads 
\begin{equation}
  \hat{H}_{\lambda}^{SDM} =  \Big( \sum_{\mathbf{k}} \sum_{\sigma}  (\Phi_{\sigma}(N, M)  \epsilon_{\mathbf{k} \sigma}   -\sigma h_a  - \mu )\hat{n}_{\mathbf{k} \sigma} \Big) - \lambda_{1}(\hat{N} - N) - \lambda_{2}(\hat{M} - M),
\label{sdm Hamiltonian}
\end{equation}
where $(-\sum_{\mathbf{k}} \sigma h_a \hat{n}_{\mathbf{k} \sigma})$ is the Zeeman term, and $h_a$ is the reduced  applied magnetic field.
The present case is, in fact, quite similar to the previous one, the band-narrowing factors $\Phi_{\sigma}(N, M)$ are  also derived using the Gutzwiller approach. Then analogously, by making use of Eq. (\ref{mean value of H GMF with lambda, d/dA}), either with $A_w = N $ or $A_w = M $,  we  obtain respectively
\begin{eqnarray}
\lambda_{1} &=& -\Big \langle   \sum_{\mathbf{k}} \sum_{\sigma} \frac{\partial \Phi_{\sigma}(N, M)}{\partial N}  \epsilon_{\mathbf{k} \sigma}  \hat{n}_{\mathbf{k}\sigma} \Big \rangle_{\lambda},
\label{lambda N zero}
\end{eqnarray}
\begin{eqnarray}
\lambda_{2} &=& -\Big \langle   \sum_{\mathbf{k}} \sum_{\sigma} \frac{\partial \Phi_{\sigma}(N, M)}{\partial M}  \epsilon_{\mathbf{k} \sigma}  \hat{n}_{\mathbf{k}\sigma} \Big \rangle_{\lambda}.
\label{lambda M zero}
\end{eqnarray}
Regardless of the detailed  analytical form of $\Phi_{\sigma}(N, M)$, we  multiply (\ref{lambda M zero}) by
%\begin{equation} -\frac{\partial \Phi_{\downarrow}(N, M)}{\partial N} \Big(\frac{\partial \Phi_{\downarrow}(N, M)}{\partial M}\Big)^{-1} \label{} \end{equation} 
\begin{equation}
-\frac{\partial}{\partial N} \Phi_{\downarrow}(N, M) \Big(\frac{\partial }{\partial M} \Phi_{\downarrow}(N, M)\Big)^{-1} \nonumber
\end{equation}
and add it to (\ref{lambda N zero}). The resulting r.h.s is then proportional to $ \big  \langle E_{kin, \uparrow}  \big \rangle_{\lambda} =  \big  \langle \sum_{\mathbf{k}}  \epsilon_{\mathbf{k} \uparrow}  \hat{n}_{\mathbf{k}\uparrow}  \big \rangle_{\lambda} $, the average band energy of particles with $z$-spin component $\sigma = \uparrow$. If we put $\lambda_{1}= \lambda_{2}=0$, all the averages become identical to those of the standard treatment. Consequently, for (\ref{lambda N zero}) and (\ref{lambda M zero}) to hold in such a case we have to require that $ \big  \langle E_{kin, \uparrow}  \big \rangle= 0$, for any admissible value of $N, M$, and functional dependence of $\Phi_{\sigma}(N, M)$ on its variables,  which is a trivial and inconsistent result. Obviously, the same  holds  for $\sigma = \downarrow$.

The numerical studies of this model and its extensions are carried out recently, \cite{Kaczmarczyk Spalek}. 

%%%%%%%%%  C O N C L U S I O N S 
\section{Summary and conclusions\label{summary}}
In this paper we have presented the method,  based on the maximum entropy principle,  of constructing the grand-canonical density operator for a wide class of  mean-field Hamiltonians. We have shown, that such a density operator is not, in general, obtained by replacing the exact Hamiltonian by its mean-field counterpart in the grand canonical density operator. Usually, some extra terms, which may be interpreted as a kind of source terms or molecular fields, must be added to the mean-field Hamiltonian. This modification also ensures consistency of basic thermodynamic relations, which is not guaranteed for self-consistent methods which are not based on the variational principle. 

It is also shown, that although for the Hartree-Fock type of mean-field Hamiltonians application of our method is not necessary, there are also some important examples, for which it does not reduce to the standard approach. The renormalized  mean-field   t-J Hamiltonian is  such case.

If there is $M$ expectation  values appearing in a mean-field Hamiltonian that are to be determined, the method presented here results in  a system of $2M$ equations. This is in contrast to the standard treatment, where only $M$ such equations appear.% A detailed numerical analysis is required to validate our results on concrete applications.

The problem of the application of maximum entropy inference in the mean-field case was studied also in a series of the papers \cite{Argentynczycy}. However, the presented point of view and the results obtained  differ essentially from ours.

Also, the problem partly related to that discussed in the present paper, namely the construction of  variational principles suited for optimization of physical quantities of interest, is examined in an interesting article \cite{Balian}. This paper contains  a detailed  analysis of  the problem, as well as covers a large area of the subject (for example the authors discuss time-dependent formalism).

%\subsection*{Appendix A. Hellmann-Feynman theorem \label{Feynman}}
%When deriving (\ref{derivative A s}) and  (\ref{derivative lambda s}) we used the following. Suppose that some operator $\hat{O}$, as well as its  eigenstate $|i\rangle$ (i.e. $\hat{O}|i\rangle = {O}_{i}|i\rangle$), depend on a parameter $x$. Then 

%\begin{eqnarray}
%\frac{\partial O_i} {\partial x} & = & \frac{\partial \langle i(x)|\hat{O}(x) |i(x)\rangle} {\partial x} =   \langle \frac{\partial i(x)}{\partial x}|\hat{O}(x) |i(x)\rangle +       \langle i(x)|\hat{O}(x) |\frac{\partial i(x)}{\partial x}\rangle +   \nonumber \\
%&+& \langle i(x)|\frac{\partial \hat{O}(x) }{\partial x}|  i(x) \rangle =  O_i \frac{\partial  }{\partial x}\langle i(x)|  i(x) \rangle + \langle i(x)|\frac{\partial \hat{O}(x) }{\partial x}|  i(x) \rangle \nonumber \\ &=&  \langle i(x)|\frac{\partial \hat{O}(x) }{\partial x}|  i(x) \rangle, 
%\end{eqnarray}
%for any normalized state $| i(x) \rangle$, where $ \langle \frac{\partial i(x)}{\partial x}| =  \frac{\partial}{\partial x}(\langle  i(x)|)$.  In Eqs. (\ref{derivative A s}) and (\ref{derivative lambda s}) we have to set $\hat{O} \equiv \hat{H}_{\lambda} - \mu  \hat{N}$.

%We may also need a related result, suppose now that we evaluate trace of some operator $\hat{0}$ but not in an eigenbasis of $\hat{0}$/ its eigenbasis. Still both  operator and the basis vectors may depend on a parameter $x$. Now

\subsection*{Acknowledgments}
The authors are grateful to Krzysztof Ro\'{s}ciszewski for  his valuable comments and insightful remarks. 
Discussions with Marcin Raczkowski and Andrzej Kapanowski are also warmly acknowledged.
The authors acknowledge the Grant No. N N 202 128 736 from the Ministry of Science and Higher Education.

\vspace{0.3cm}

$\ast$e-mail: jedrak@th.if.uj.edu.pl 

$\dag$e-mail: ufspalek@if.uj.edu.pl\\

\end{document}